# A thermally grown $SiO_2$ diffusion barrier enabling high-temperature investigation of Ag-Au-Pd-Pt thin films

*Elaheh Akbarnejad[1], Aleksander Kostka[2], Advika Chesetti[2], Alan Savan[1], Georg Fritz[3], Kamen Kozhuharov[3], Matthias Karl Klein[3], Yujiao Li[2]*, Alfred Ludwig[1,2,4]**

[1]Materials Discovery and Interfaces, Institute for Materials, Ruhr University Bochum, Bochum, 44801, Germany

[2] Center for Interface-Dominated High-Performance Materials (ZGH), Ruhr University Bochum, Bochum, 44801, Germany

[3] Team Nanotec GmbH, Wilhelm-Schickard-Straße 10, 78052 Villingen-Schwenningen, Germany

[4] Research Center Future Energy Materials and Systems (RC FEMS), Ruhr University Bochum, Bochum, 44801, Germany

*Corresponding authors: alfred.ludwig@rub.de, yujiao.li@rub.de

**Abstract:**
Combinatorial processing platforms (CPPs), integrating Si microtip arrays with combinatorial thin film synthesis and atom probe tomography (APT), enable near-atomic-scale characterization of compositionally complex solid solutions (CCSSs) under diverse processing and reaction conditions, including oxidation, thermal phase stability and electrocatalytic reactions. Their application at elevated temperatures, however, can be limited when CCSS constituents such as Pd and Pt react with the Si support to form silicides. Although thermally grown $SiO_2$ has proven effective as a diffusion barrier between pure Pt and Si, its performance for multicomponent CCSS thin films is unclear. Here, using Ag-Au-Pd-Pt as a model system, we compare a 25 nm thermally grown $SiO_2$ barrier with native Si oxide during annealing using APT and transmission electron microscopy. Native Si oxide prevents detectable interfacial reactions up to 300°C, but at 400°C Pd and Pt react with Si, causing silicide formation and substantial redistribution of the film constituents. At 600°C, extensive substrate reactions disrupt the CCSS film and produce a pronounced needle-shaped silicide morphology. In contrast, thermally grown $SiO_2$ suppresses CCSS thin film-substrate reactions up to 600°C and retains the CCSS composition. The thermally grown $SiO_2$ thus extends the applicable temperature range of Si-based CPPs to at least 600°C for near-atomic-scale characterization of CCSS thin films.

## Introduction

Compositionally complex solid solution alloys (CCSS) are of interest for technical applications where a synergy between competing material properties is required [1,2]. To achieve this, atomic-scale understanding of phase evolution and stability in dependence of temperature, time, and environment, is essential for tailoring the microstructure and, ultimately, the resulting material properties [3,4]. However, given the vast compositional space associated with CCSS, accelerating their discovery and development requires combinatorial synthesis and high-throughput screening coupled with direct atomic-scale characterization to systematically assess their thermal and chemical stability under certain processing conditions [5]. The combinatorial processing platform (CPP) approach [6–9] enables rapid, atomic-scale investigation of compositionally complex thin-film materials using atom probe tomography (APT) and transmission electron microscopy (TEM) [9]. APT provides three-dimensional compositional characterization with near-atomic-scale spatial resolution and high chemical sensitivity [10–13]. In this approach, 36 identical films are sputter-deposited simultaneously on a commercially available, pre-sharpened Si microtip array (Cameca Instrument, Inc), eliminating the need for focused-ion beam (FIB) based sample preparation. The coated tips are directly analyzed by APT after synthesis or after various processing steps.

However, a persistent challenge is the high reactivity of Si with several metals, particularly transition and noble metals such as Ni, Pd, and Pt, which can form silicides at relatively low annealing temperatures [14]. Previous studies have investigated native $SiO_2$ and an $Al_2O_3$ layer deposited by atomic layer deposition (ALD), as diffusion barriers in the CPP approach using Pt as a model system. The native $SiO_2$ was effective at lower temperature, but Pt diffuses through the oxide resulting in Pt-silicide formation above 400°C, whereas a 20 nm ALD-grown $Al_2O_3$ layer extended the barrier performance up to 600°C [15]. However, despite its good barrier properties, $Al_2O_3$ complicates APT analysis due to the differences in field evaporation behavior between the oxide and the specimen, which can lead to trajectory distortions and premature tip fracture, resulting in reduced analysis yield [15]. The relatively early failure of the native $SiO_2$ as a diffusion barrier [16], leading to unwanted interfacial reactions, may be associated with its limited thickness of approximately 2 nm, motivating the investigation of a substantially thicker $SiO_2$ layer [17].

Thermal oxidation is particularly suitable for this purpose because it produces a conformal $SiO_2$ layer directly from the Si substrate while preserving the geometry of the PSM array. In our previous study, we demonstrated that 20-50 nm thermal grown $SiO_2$ layers effectively suppressed Pt-Si interdiffusion and silicide formation up to 800°C, while also delaying film dewetting [18]. The study established thermally grown $SiO_2$ as an effective diffusion barrier for Pt films in the CPP approach. However, whether comparable barrier performance can be maintained for compositionally complex, multicomponent films remains unresolved. In such materials, several elements with different affinities towards Si coexist and their redistribution during annealing may additionally modify the local conditions at the interface between film and diffusion barrier.

Ag-Au-Pd-Pt CCSS thin films provide a suitable model system for evaluating this question. Besides their use as compositionally tunable catalytic materials [19], its constituent elements exhibit contrasting behavior with Si: Ag and Au exhibit negligible solubility in Si [20–23] and do not form stable silicides, whereas Pd and Pt readily form silicides [24–26]. In our previous study, the alloy was deposited directly onto Si tips protected only by the native oxide layer [27]. However, substrate interactions became evident from 400°C, as Pd and Pt diffused through the native oxide and reacted with Si to form silicides, thereby interfering with the intrinsic phase evolution of the CCSS. Therefore, this limitation necessitates the use of a more effective diffusion barrier to investigate the phase stability of Ag-Au-Pd-Pt at elevated temperatures.

In the present study, we evaluate a 25 nm thermally grown $SiO_2$ layer as a diffusion barrier for Ag-Au-Pd-Pt CCSS films deposited on Si-based CPPs. Identical films on microtip arrays containing either native oxide or thermally grown $SiO_2$ are annealed at temperatures up to 600°C and characterized by APT and complementary TEM. By comparing elemental redistribution, silicide formation, thin film composition and morphology between the two interfaces, we determine the temperature-dependent effectiveness of the diffusion barriers and establish the extent to which thermal $SiO_2$ suppresses substrate-induced modification of the CCSS thin film.

## Materials and methods

Commercial pre-sharpened microtips (PSM) chips comprising 36 Si tips (Microtip™ arrays, CAMECA Instruments, Inc.) were used for the combinatorial processing investigation. PSMs were subjected to thermal oxidation in a horizontal furnace (CPS382-II, Centrotherm) in $O_2$ atmosphere at 1080°C for 4 min to form thermal $SiO_2$ diffusion barriers with a nominal thickness of 25 nm. For direct comparison, another PSM with only native oxide was used.

The PSM chips were placed on a micromachined carrier wafer with a cavity slightly larger than the PSM size to fix the 3×7 $mm^2$ coupon during pumping and deposition. A magnetron sputter deposition system (DCA Instruments, Finland) was used for CCSS film deposition. The 100 mm diameter elemental targets of Ag (99.99%, K.J. Lesker, DC), Au (99.99% Sindlhauser, RF), Pd (99.99% Kaistar, RF) and Pt (99.99% MaTecK, DC) were used to deposit films on the PSMs. The base vacuum was $1.5x10^{-6}$ Pa and the deposition was carried out under Ar (99.9999%) pressure of 0.67 Pa with the substrate rotating at 10 rpm and 25°C without additional intentional heating, yielding a uniform composition with a thickness of about 50 nm.

The elemental composition of the deposited film was measured on an adjacent separate substrate of thermally oxidized Si by energy dispersive X-ray spectroscopy (EDS) in a scanning electron microscope (SEM, JEOL JSM-5800) with quantitative analysis performed using Oxford Instruments INCA software.

APT analysis was performed using a LEAP 5000 XR (CAMECA Instruments, Inc.) in laser mode at 65 K. A laser energy of 80 pJ and a pulse frequency of 125 kHz were used, with a detection rate of 0.004 atoms per pulse. The 3D datasets were reconstructed and analyzed using IVAS 3.8.16 software.

Following established procedures [28], APT tips were transferred from the CPPs to TEM grids using an FIB system (FEI Helios G4 CX) operated at 30 kV. To enable electron transmission for high-quality TEM imaging and reliable crystal structure analysis, the tips were transferred from the CPPs to a TEM copper grid and subsequently thinned down to typically 50 - 70 nm using procedures like those used for TEM lamella preparation. To reduce beam-induced damage, the ion beam voltage was reduced to 16 kV and during the final thinning step to 8 kV. The thinning process was performed starting from the mid-section of the APT tips, ensuring uniform reduction in thickness. This method enables near-atomic resolution TEM imaging and analysis, allowing detailed observation of interfacial reactions.

An aberration-corrected TEM (JEOL JEM-ARM200F) operated at 200 kV and equipped with two JEOL EDS detectors was used to characterize the composition and microstructure of the CPPs. EDS results were normalized to 100% based on detector counts without corrections for thickness or absorption. Carbon, gallium and copper were excluded from quantification as sample preparation artifacts. Phase identification was performed using the Crystallographic Tool Box (CrysTBox) [29] with reference crystal data from the International Crystal Structure Database (ICSD); Pt: 41525, PdSi: 15016, $Pd_2Si$: 43209, Ag: 180878.

To evaluate the effectiveness of the $SiO_2$ diffusion barrier at elevated temperatures, the CPPs coated with Ag-Au-Pd-Pt films, one protected only by native oxide, referred to as CCSS/Si CPP, and the other with a 25 nm thermally grown $SiO_2$ diffusion barrier, referred to as CCSS/$SiO_2$ CPP, were annealed at temperatures up to 600°C. Annealing was performed in a tubular furnace under a vacuum of $8.5\times10^{-5}$ Pa, starting at 100°C for 1 h. After each annealing step, the furnace was cooled to room temperature before the samples were removed for APT and TEM characterization to assess elemental redistribution, film-substrate interdiffusion, and interfacial reactions. This process was repeated for subsequent annealing steps up to 600°C.

## Results and discussion

Figure 1 shows TEM bright field (BF) images of an APT tip from the CCSS/$SiO_2$ CPP for the as-deposited state. For the TEM sample preparation, the APT tip was covered with a 50 nm protective carbon layer before exposure to the ion beam. The thick protection carbon layer visible over the APT tip in Fig. 1a was obtained with the ion beam and is necessary to ensure good thinning conditions for TEM transparency.

Figure 1b reveals the nanocrystalline grain structure of the film in the tip. The average grain size is approximately 15 nm. The targeted layer thickness of the CCSS (about 50 nm) and the $SiO_2$ diffusion barrier (about 25 nm) were achieved, as shown in Fig. 1c. The selected area diffraction pattern acquired from the CCSS region, indicated by the white arrow in Fig 1b, confirms a single-phase face-centered cubic (fcc) structure.

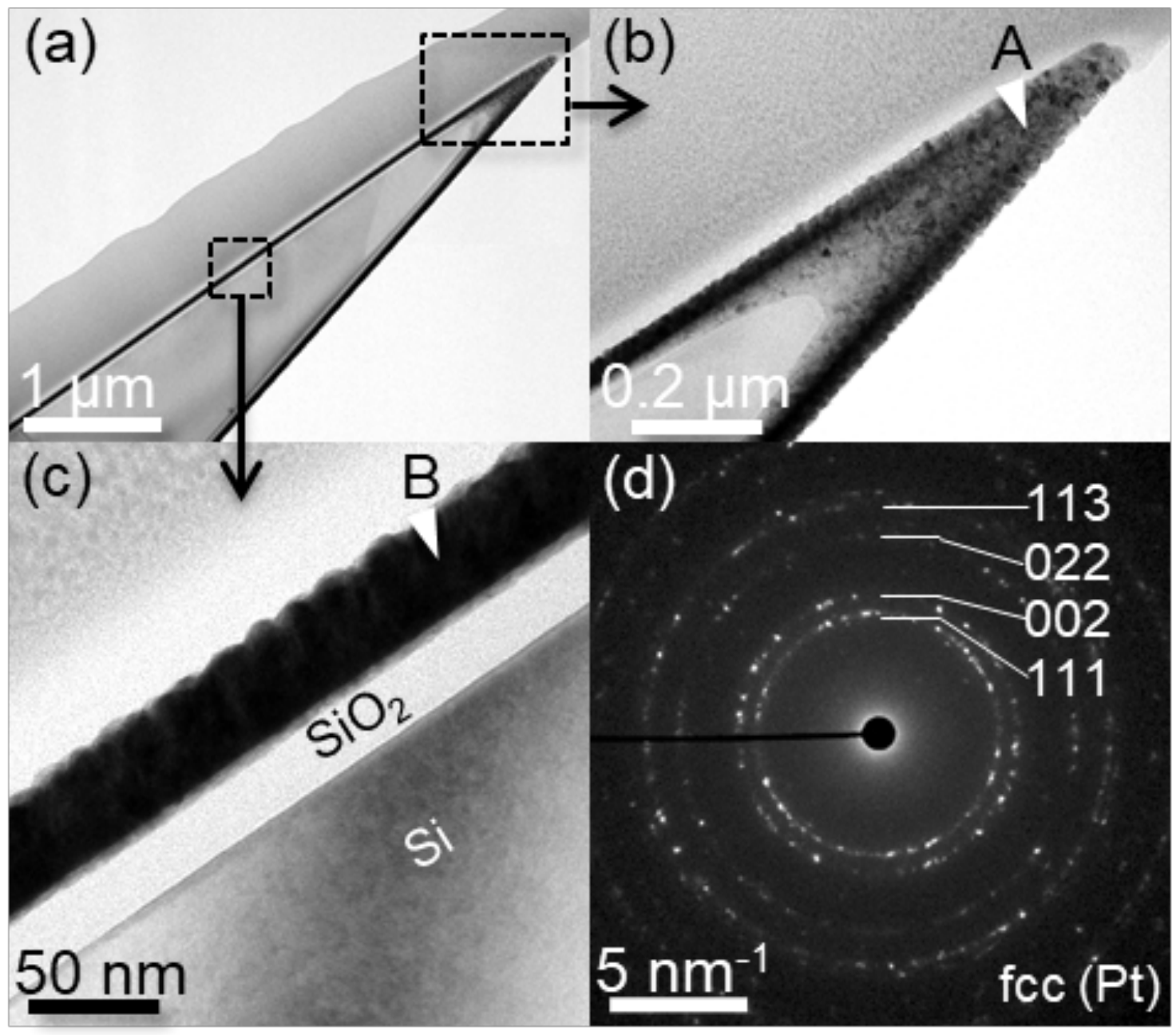

Figure 1: Results of TEM analyses of the Ag-Au-Pd-Pt CPP in the as-deposited state. (a-c) BF images of the film's microstructure and the interfaces between the Si APT tip. Capital letters refer to the EDS measurements shown in Table 1. (d) The selected area diffraction (SAD) pattern was collected from the APT tip shown in (b). The indexed planes refer to fcc (Pt).

Since annealing below 300°C showed no detectable changes in elemental distribution, phase structure, or interdiffusion between the film and Si for both CPPs, these results are omitted here for brevity. Figure 2 presents the APT results of the CCSS/Si CPP following annealing at 300°C for 1 h and subsequently at 400°C for 1 h.

After annealing at 300°C (Figs. 2a and 2b), the CCSS coating retained its columnar grain structure as indicated by the high number density of atoms at the grain boundaries in the elemental distribution maps in Fig. 2a. No evidence of chemical inhomogeneity is observed within the CCSS thin film in the 1D composition profile shown in Fig. 2b. The average composition of the CCSS film based on the 1D composition profile (Fig. 2b) is approximately 24 at. % Ag, 18 at. % Au, 8 at. % Pt, and 50 at. % Pd. The native oxide layer surrounding the Si core remained present, and no significant interdiffusion between the coating and Si was detected. Localized Pd enrichment was observed only in isolated regions, as indicated by the blue arrow in Fig. 2a, with no evidence of a continuous interfacial reaction. Overall, native oxide remained effective in suppressing detectable interaction between the CCSS film and the Si core at 300°C.

A clear change is observed after subsequent annealing at 400°C (Fig. 2c), indicating the onset failure of the native oxide barrier. The elemental distribution is no longer homogeneous: Pt- and Pd-enriched regions accompanied by Ag- and Au-depleted regions, indicate decomposition of the CCSS. In addition, Pd and Pt are detected within the Si tip, as shown in the proxigram in Fig. 2d, demonstrating interdiffusion across the native oxide barrier The composition near the film-substrate interface deviates significantly from the nominal composition of the CCSS film observed at 300°C (Fig. 2b), which is also corroborated by the chemical inhomogeneity observed in the elemental distribution maps in Fig. 2c. The elemental contents of the CCSS film before decomposition are represented by the dash-dotted lines in Fig. 2d for comparison. These results demonstrate that interdiffusion with the Si substrate substantially alters the composition of the original CCSS film. Consequently, the phase decomposition observed after annealing at 400°C is classified as substrate-induced decomposition rather than intrinsic phase evolution of the nominal Ag-Au-Pd-Pt composition.

Nevertheless, quantifying the compositions of the decomposed constituents is useful in comparison with the intrinsic phase decomposition observed later in the presence of the thermal $SiO_2$ barrier. In particular, the Pt-Pd-rich and Ag-Au-rich regions in the thin film away from the native $SiO_2$ barrier, as indicated by ROI 2 in Fig. 2c, were analyzed to determine the local compositions resulting from the strong depletion of Pd and Pt from the remaining film due to its interaction with the Si core. The corresponding 1D composition profile of ROI 2 is shown in Fig. 2e. The Ag-Au rich Zone I (Fig. 2e) comprises ≈ 52 at. % Ag, 28 at. % Au, 15 at. % Pd, and 5 at. % Pt, while the Pt-Pd rich Zone II has ≈ 38 at. % Pd, 9 at. % Pt, 31 at. % Ag, and 19 at. % Au. Additionally, the Pd and Pt concentrations after interaction with the Si are ≈ 41 at. % and 22 at. % respectively with 35 at. % Si.  Based on the composition of the intermixed Pd, Pt, and Si (Fig. 2d), and the (Pd-Pt): Si ratio of 1.8, it is likely that a silicide, $(PdPt)_2Si$, formed. Our previous work also showed a ternary Pd-Pt-Si silicide with a crystal structure similar to $Pd_2Si$, belonging to the $P\bar{6}m2$ space group [27].

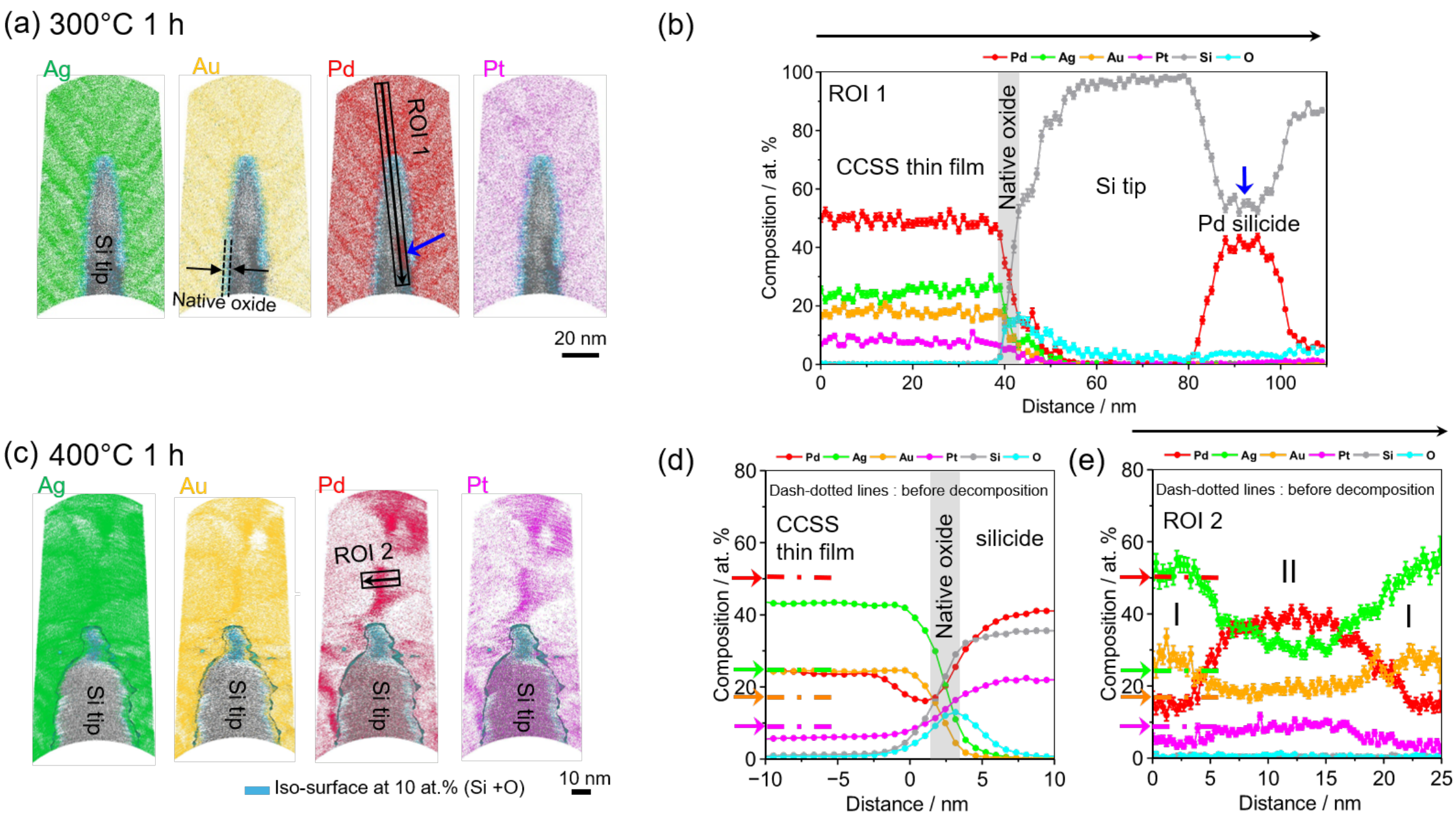

Figure 2: Results of the APT analysis of the CCSS film with native oxide $SiO_2$ diffusion barrier after annealing. (a) Elemental distributions of Ag (green), Au (yellow), Pd (red) and Pt (pink) after annealing at 300°C for 1 h. The black rectangle marks the region of interest (ROI 1) selected for quantitative composition analysis. The black arrow indicates the direction along which the 1D composition profiles are plotted. (b) 1D compositional profile of ROI 1, (c) Elemental distribution maps in the state after annealing 400°C for 1 h with the cyan-colored iso-concentration surface plotted at 10 at.% (Si+O), highlighting the Si tip. (d) Proxigram plotted for the 10 at. % (Si+O) iso-concentration surface shown in (c). The black rectangle in the Pd distribution map in (c) marks the region of interest (ROI 2), with the arrow indicating the direction along which the 1D composition profile is considered. (e) 1D composition profile of ROI 2 with two distinct regions denoted as Zones I and II in the graph. The dash-dotted lines are marked for reference in (d) and (e) to represent the elemental concentrations in the CCSS before decomposition.

These results demonstrate that once the native oxide barrier fails, Pd and Pt are selectively depleted from the CCSS film through interdiffusion and silicide formation with the Si core. The resulting compositional changes therefore modify the phase evolution of the original film, making it impossible to distinguish substrate-induced decomposition from the intrinsic thermal evolution of the nominal Ag-Au-Pd-Pt CCSS. A more effective diffusion barrier is therefore required to suppress substrate-induced reactions and enable the intrinsic thermal phase evolution of the CCSS to be investigated.

Figure 3 focuses on the role of the thermally grown 25 nm $SiO_2$ diffusion barrier in preventing diffusion and intermixing. Fig. 3a shows the elemental distribution maps after annealing the CCSS/$SiO_2$ CPP at 300°C for 1 h, and Fig. 3c shows the corresponding results after annealing at 600°C for 1 h. At 300°C (Fig. 3a), no phase decomposition or interdiffusion is detected. Figure 3b presents a quantitative compositional analysis for a region of interest (ROI 1), which spans the CCSS thin film, the thermally grown $SiO_2$ layer, and the Si core. No detectable intermixing occurs across the CCSS/$SiO_2$/Si interfaces.

After annealing at 600°C (Fig. 3c), a Pt-enriched phase forms within the CCSS thin film. Two regions of interest: ROI 2 and ROI 3, were selected to quantify the composition of the Pt-enriched phase and to assess the CCSS/$SiO_2$ interface, respectively. Zone I in the 1D composition profile from ROI 2 (Fig. 3d) contains ≈ 22 at. % Pt, 53 at. % Pd, 8 at. % Ag, and 9 at. % Au compared with the matrix composition in Zone II of ≈ 4 at. % Pt, 47 at. % Pd, 30 at. % Ag, and 17 at. % Au. Importantly, the Pt-enriched phase forms without detectable interdiffusion between the CCSS film and the underlying Si core. In contrast to the substrate-induced decomposition observed for the native-oxide CPP at 400°C, the phase separation observed here is therefore attributed to the intrinsic thermal evolution of the Ag-Au-Pd-Pt CCSS.

The 1D composition profile from ROI 3 (Fig. 3e) shows no detectable Pd or Pt diffusion into the Si core, in contrast to the CCSS/Si CPP annealed at 400°C for 1 h (Fig. 2d). The CCSS composition remains close to the nominal composition across the film. These results demonstrate the thermally grown $SiO_2$ diffusion barrier effectively separates the CCSS film from the Si core up to 600°C, allowing its intrinsic phase evolution to be investigated without detectable substrate interference.

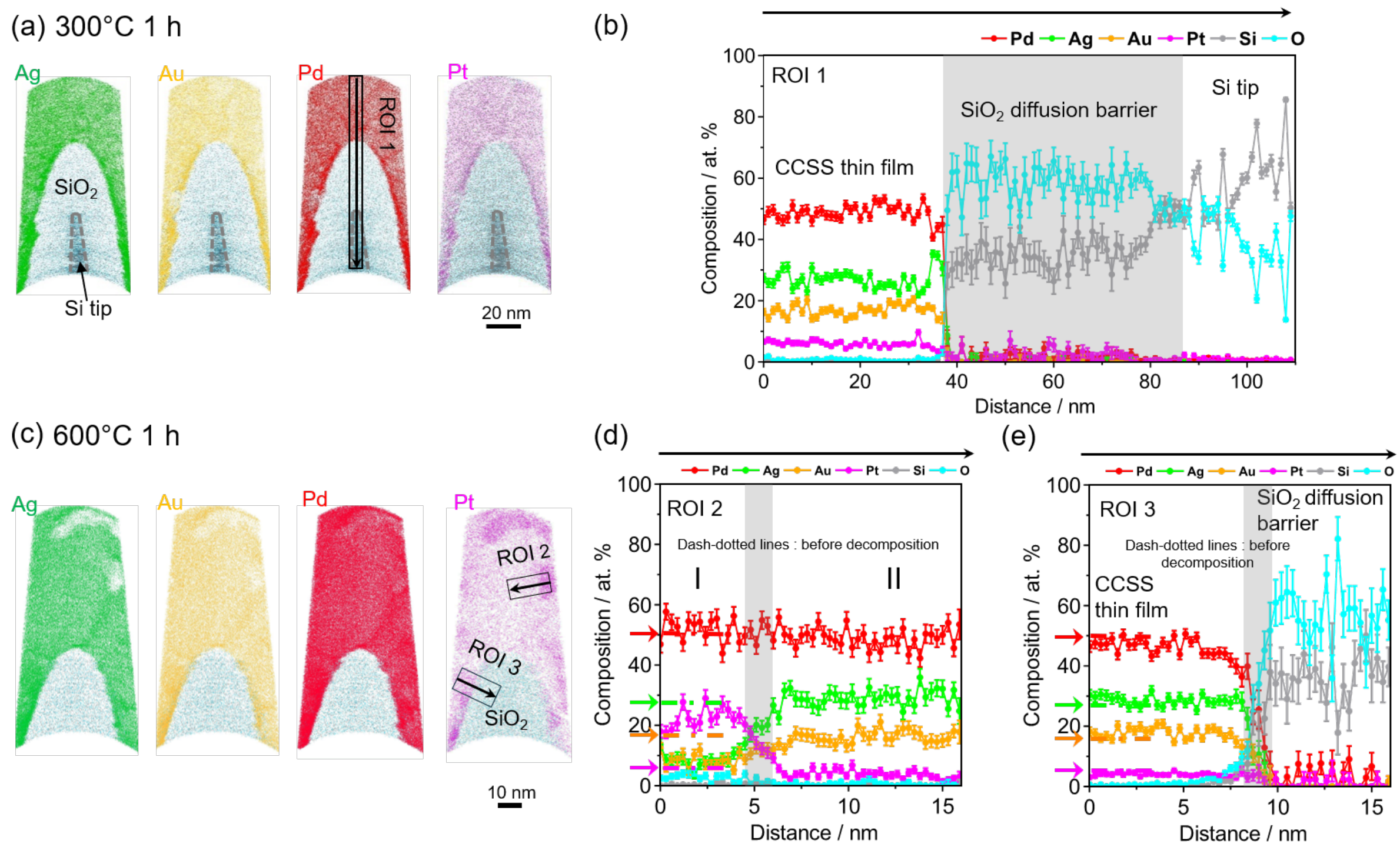

Figure 3: Results of the APT analysis of the CCSS film with a 25 nm thermal $SiO_2$ diffusion barrier after annealing at (a) 300°C and (c) 600°C for 1 h. Panels (a) and (c) show the elemental distributions of Ag (green), Au (yellow), Pd (red), and Pt (pink). In (a), the black rectangle marks ROI 1 (5×7×110 nm$^3$) across the CCSS film, $SiO_2$ barrier, and Si tip; the corresponding 1D composition profile is shown in (b). In (c), ROI 2 (9×5×16 nm$^3$) and ROI 3 (7×5×16 nm$^3$) identify regions for localized analysis. Their 1D composition profiles are shown in (d) and (e), respectively, highlighting the Pt-rich phase (ROI 2) and the transition from the thin film into the thermal $SiO_2$ (ROI 3). The distinct regions in the 1D composition profile in (d) are denoted by Zones I and II. The dash-dotted lines are marked for reference in (d) and (e) to represent the elemental concentrations in the CCSS before decomposition.

To verify whether any trace amount of Ag, Au, Pd, or Pt is present in the $SiO_2$ layer and vice-versa for the CCSS/$SiO_2$ CPP, detailed mass spectra analyses were conducted separately for two ROIs (Fig 4a): ROI 1 from the CCSS film (blue box), and ROI 2 from the $SiO_2$ layer (black box). The tables provided as insets in Fig. 4a represent the nominal composition of the CCSS thin film and $SiO_2$ layer. Based on the mass spectrum from ROI 1 in Fig. 4b, there is no indication of distinct $Si^+$, $Si^{2+}$, $SiO^+$, $SiO^{2+}$ and $SiO_2^+$ peaks. In the case of the mass spectrum from ROI 2 in Fig. 4c, the $Si^+$, $Si^{2+}$, $SiO^+$, $SiO^{2+}$ and $SiO_2^+$ peaks are observed, however no distinct peaks of Ag, Au, Pd, and Pt at higher charge states are present in the thermal $SiO_2$ region. Furthermore, based on the overall mass spectrum of ROI 2 in Fig. 4d, the $Pd^+$, $Pt^+$, $Ag^+$, and $Au^+$ peaks which occur in the range of 105 to 197 Da are also not observed, confirming that there is no interdiffusion of the elements from the CCSS thin film into the $SiO_2$ layer. Additionally, while the O peak has a high intensity in the mass spectrum for ROI 2, peaks associated with O and hydrocarbons are observed in the case of ROI 1 as well. However, based on the ranged mass spectrum of ROI 1, less than 2 at. % of O, H, and C are present in the CCSS thin film. The trace amounts of O, H, and C can be attributed to residual contamination from the sputter system, the APT chamber, and/or exposure during sample transfer.

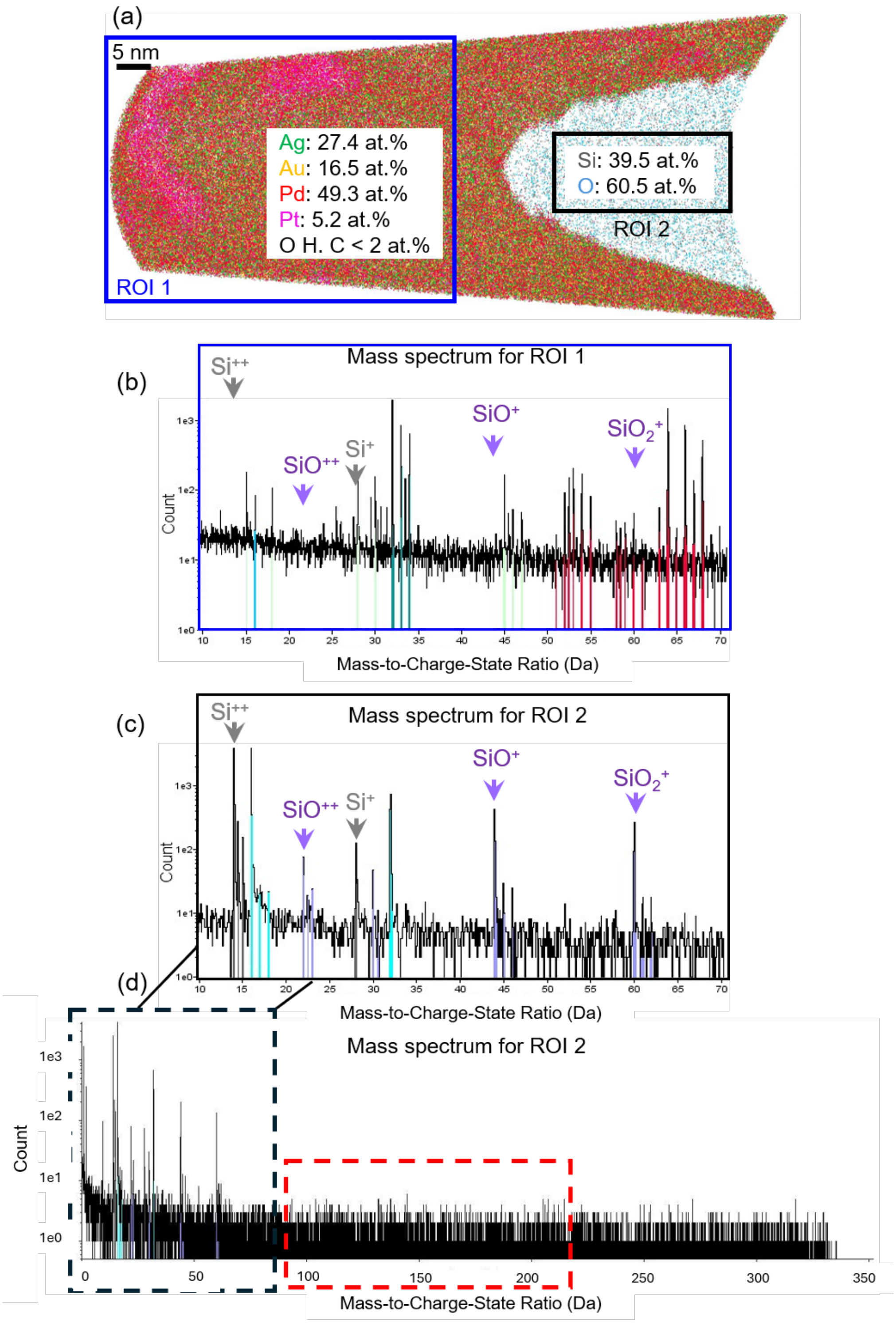

Figure 4: APT analysis of trace elements to assess potential interactions between the CCSS thin film and the Si after annealing at 600°C for 1 h. (a) Overall elemental distribution of Ag, Au, Pd and Pt in the CCSS film sputtered on the $SiO_2$/Si substrate part. The inset tables show the average compositions of the corresponding ROIs, indicated by the blue box (ROI 1: 42×40×50 $nm^3$) and the black cylinder (ROI 2: cylinder diameter of 12.65 nm, Z: 25 nm). The ROIs were selected to separately analyze the mass spectra of the CCSS layer and the $SiO_2$ diffusion-barrier region. (b) APT mass spectrum of ROI 1 and (c) APT mass spectrum of ROI 2, plotted over a mass-to-charge ratio range of 10-70 Da. (d) Mass spectrum from ROI 2 plotted up to 350 Da where the black dashed box represents the range shown in (c) and the red dashed box represents the range where $Pd^+$, $Pt^+$, $Ag^+$, and $Au^+$ peaks are expected to be observed.

TEM analysis (Fig. 5) provides a comparative assessment of the Ag-Au-Pd-Pt CCSS film with native oxide and with thermal $SiO_2$ diffusion barrier after annealing at 600°C for 1 h. In the absence of a thermal $SiO_2$ diffusion barrier, the native oxide is insufficient to prevent the formation of second-phase compounds. The SEM image (Fig. 5a) reveals a surface decorated with needle-like crystals. Following the TEM analysis (Fig. 5b), it shows the formation of two new phases: (i) an Ag-Au-rich phase highlighted by E, and (ii) PdSi silicide highlighted by F (with the corresponding SAD shown in Fig. 5e and 5f and Table 1). After the formation of the new phases, the shank part of the CPP has a thickness of about 18 nm (50 nm after deposition) and is decorated with Pd-silicide crystals extending out of the surface (bottom-left corner of Fig. 5c and 5d). This is confirmed by the EDS map showing an overlay of Pd, Ag, and Si in Fig. 5d, where the Pd-silicides extending out of the surface correspond to the region marked as F, discerned to be $(Pd\text{-}Pt)_2Si$ based on its measured composition in Table 1.

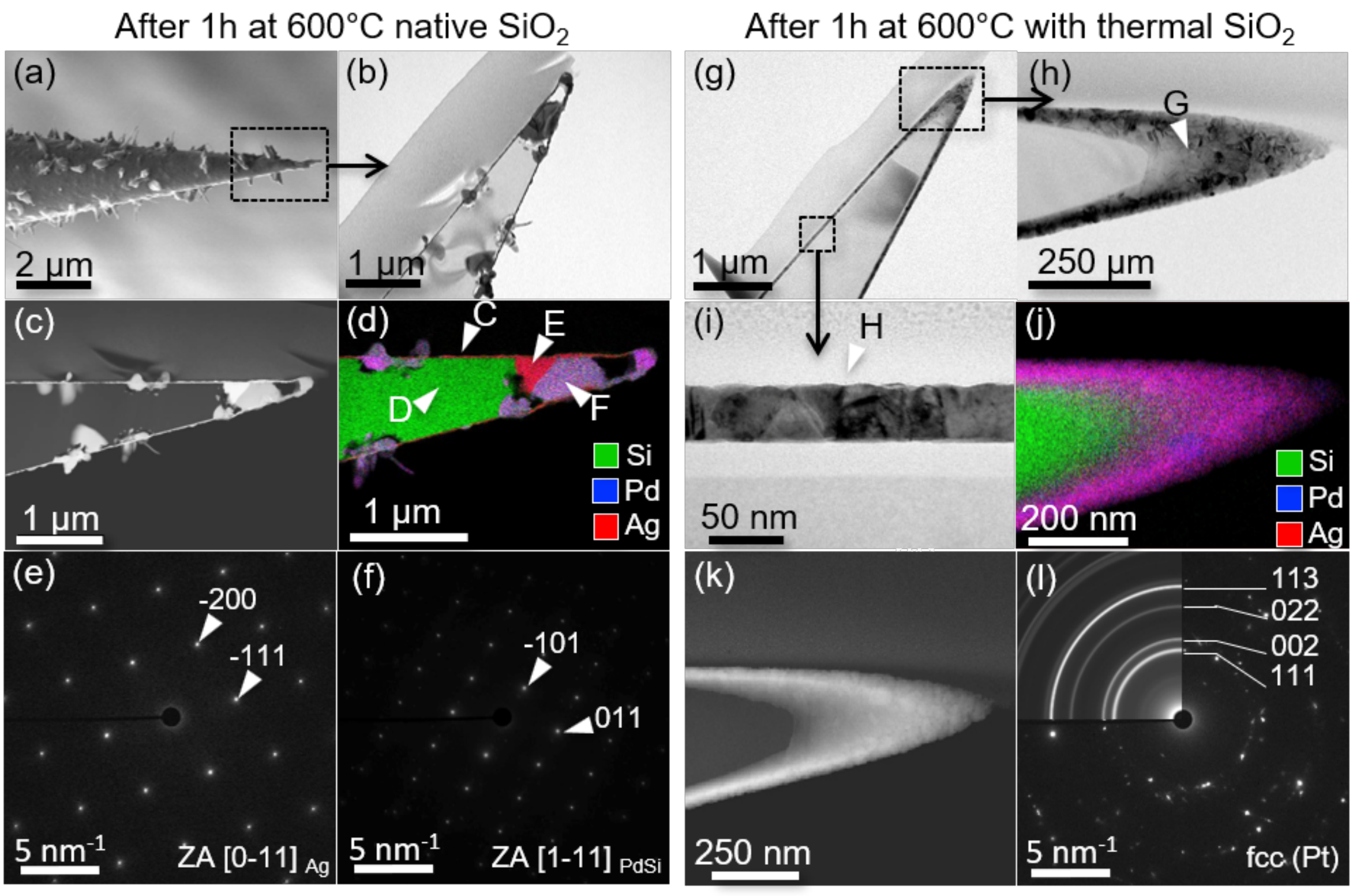

Figure 5: Results of TEM analyses of the Ag-Au-Pd-Pt CCSS/CPP with native Si oxide and with thermal $SiO_2$ after annealing at 600°C for 1 h. (a) SEM image showing the morphology of the annealed APT tip with the native oxide. The highlighted region was used for the preparation of the TEM sample shown in (b). (c) ADF STEM image of the thinned tip and (d) the corresponding EDS map showing an overlay of Ag, Pd and Si. Capital letters indicate the regions used for quantitative EDS measurements summarized in Table 1. (e) Indexed SAD pattern from the region highlighted as E in the figure (d). (f) Indexed SAD pattern from the region highlighted as F in (d). (g, h, i) BF TEM images of the APT tip with the 25 nm thermal $SiO_2$ diffusion barrier, revealing minor changes after annealing. (j) EDS map showing an overlay of Ag, Pd and Si, and (k) the corresponding ADF STEM image. (l) SAD pattern obtained from the G region marked as G in (h). The insert represents an indexed rotational averaged pattern (for clarity).

In contrast, Fig. 5g-i illustrates the microstructure where, during the annealing (1 h at 600°C), the Ag-Au-Pd-Pt CPP stays well separated from the Si by the thermal $SiO_2$ diffusion barrier, with no evident film-substrate interaction and evidenced by the overlay EDS image (Fig. 5j). The corresponding STEM image of the region considered for EDS is shown in Fig. 5k. During annealing, the average grain size of the CCSS increases from approximately 15 nm in the as-deposited state to about 60 nm, while the thicknesses of the CCSS layer and the $SiO_2$ diffusion barrier remained essentially unchanged at 50 nm and 25 nm, respectively. The SAD pattern in Fig. 5l, acquired from the region corresponding to position G in Fig. 5h, shows only a fcc structure, in contrast to the silicide formation observed in the absence of an effective $SiO_2$ diffusion barrier.

The STEM-EDS measured local compositions for both CCSS/Si (native $SiO_2$) and CCSS/$SiO_2$ (thermally grown $SiO_2$) CPPs annealed at 600°C are summarized in Table 1. The Ag-Au-Pd-Pt solid solution composition from the as-deposited condition of CCSS/$SiO_2$ (regions A and B from Fig. 1b and 1c respectively) is also provided in Table 1, while that of the CCSS/Si can be found in [27]. The average composition measured from the apex and shank regions of the CCSS/$SiO_2$ CPP at 600°C (regions G and H from Fig. 5h and 5i respectively) are observed to be close to the Ag-Au-Pd-Pt solid solution composition in the as-deposited state, confirming that there is no significant depletion of Pd and Pt from the matrix, in contrast to the CCSS/Si CPP where phase decomposition in the CCSS was observed due to Pd and Pt interaction with the Si core. The compositions of the Pd-Pt silicide and Ag-Au rich phase in CCSS/Si CPP at 600°C (regions C, E, and F from Fig. 5d) are summarized in Table 1. Therefore, the Pt enriched region observed in CCSS/$SiO_2$ CPP from APT measurements (Fig. 3c, 3d) is intrinsic to the CCSS thin film and does not result from interactions with the Si core. The Pt-enriched phase was not observed with STEM-EDS measurements likely due to its low volume fraction, and length-scale.

Table 1. Chemical composition (at. %) of the areas highlighted in Figure 5 for the annealed state at 600°C. EDS quantification is standardless and normalized to 100%. Positions are highlighted by white arrows.

| Condition | Annealing | Origin | Marker | O | Si | Pd | Ag | Pt | Au | Region |
|---|---|---|---|---|---|---|---|---|---|---|
| with thermal $SiO_2$ | as-deposited | Fig.1b | A-apex | - | - | 44.7 | 32.2 | 6.3 | 16.8 | Thin film |
| | | Fig.1c | B-shank | - | - | 44.8 | 32.5 | 6.1 | 16.6 | Thin film |
| with native $SiO_2$ | 600°C | Fig.5b | C-shank | - | - | 3.2 | 58.9 | 0.5 | 37.3 | Ag-Au rich |
| | | | D-Si | 2.8 | 97.2 | - | - | - | - | Si core |
| | | | E-apex | - | - | 2.9 | 61.0 | - | 36.1 | Ag-Au rich |
| | | | F-apex | 3.1 | 40.8 | 44.8 | - | 10.4 | 0.9 | Pd-Pt silicide |
| with thermal $SiO_2$ | | Fig.5f | G-apex | - | - | 44.6 | 32.9 | 5.1 | 17.4 | Thin film (composition preserved) |
| | | Fig.5g | H-shank | - | - | 45.4 | 31.4 | 6.3 | 16.9 | Thin film (composition preserved) |

## Summary

This study establishes a thermally grown $SiO_2$ layer as an effective diffusion barrier for extending the operating temperature range of Si-based CPPs. Native Si oxide prevents detectable interaction between the Ag-Au-Pd-Pt CCSS film and Si up to 300°C but becomes ineffective at higher temperatures. At 400°C, Pd and Pt diffuse towards the Si substrate and form silicides, depleting these elements from the remaining thin film and substantially changing its composition. At 600°C, extensive substrate reactions disrupt the original film and produce a pronounced needle-shaped silicide morphology. Consequently, the resulting film composition is strongly influenced by substrate-induced reactions.

With the 25 nm thermally grown $SiO_2$ layer, interdiffusion and substrate reactions are suppressed up to 600°C, and the overall Ag-Au-Pd-Pt film composition is largely retained. The Pt-rich phase observed at 600°C can therefore be attributed to changes occurring within the CCSS film rather than to compositional alteration caused by interaction with the Si substrate.

These results demonstrate that controlling the film-substrate interface is essential when Si-based CPPs are applied to multicomponent materials containing silicide-forming elements. Thermally grown $SiO_2$ provides a robust barrier against such reactions and broadens the temperature range over which the composition and phase constitution of CCSS thin films can be reliably characterized by APT and TEM.

## Acknowledgments

This project was supported by the Federal Ministry for Economic Affairs and Climate Action (BMWK) based on a decision by the German Bundestag; Zentrales Innovationsprogramm Mittelstand (ZIM, Central Innovation Program for Small and Medium-sized Enterprises), project numbers KK5380601ZG1 (Team Nanotec GmbH) and KK5116305ZG1 (Ruhr University Bochum). Furthermore, the RUB authors acknowledge funding by the Deutsche Forschungsgemeinschaft (DFG, German Research Foundation), SFB 1625, project-ID 506711657, subprojects A01-Ludwig, A02-Ludwig, S and B01. The ZGH at Ruhr University Bochum is acknowledged for using its facilities.

# References

[1] Z. Li, K.G. Pradeep, Y. Deng, D. Raabe, C.C. Tasan, *Metastable high-entropy dual-phase alloys overcome the strength-ductility trade-off*, Nature 534 (2016) 227–230, 10.1038/nature17981.

[2] B. Gludovatz, A. Hohenwarter, K.V.S. Thurston, H. Bei, Z. Wu, E.P. George, R.O. Ritchie, *Exceptional damage-tolerance of a medium-entropy alloy CrCoNi at cryogenic temperatures*, Nature Communications 7 (2016) 10602, 10.1038/ncomms10602.

[3] Y.-C. Qin, F.-Q. Wang, X.-M. Wang, M.-W. Wang, W.-L. Zhang, W.-K. An, X.-P. Wang et al., *Noble metal-based high-entropy alloys as advanced electrocatalysts for energy conversion*, Rare Metals 40 (2021) 2354–2368, 10.1007/s12598-021-01727-y.

[4] G.A. Ochsner, R. Mizuochi, L. Wang, Z. Wang, *Electronic and Structural Determinants of Activity and Stability in Noble Metal OER Catalysis*, ChemCatChem 18 (2026) e01869, 10.1002/cctc.202501869.

[5] A. Saksena, Y.-C. Chien, K. Chang, P. Kümmerl, M. Hans, B. Völker, J.M. Schneider, *Metastable phase formation of Pt-X (X = Ir, Au) thin films*, Scientific Reports 8 (2018) 10198, 10.1038/s41598-018-28452-4.

[6] A. Ludwig, *Discovery of new materials using combinatorial synthesis and high-throughput characterization of thin-film materials libraries combined with computational methods*, npj Computational Materials 5 (2019) 70, 10.1038/s41524-019-0205-0.

[7] Y.J. Li, A. Kostka, A. Savan, A. Ludwig, *Atomic-scale investigation of fast oxidation kinetics of nanocrystalline CrMnFeCoNi thin films*, Journal of Alloys and Compounds 766 (2018) 1080–1085, 10.1016/j.jallcom.2018.07.048.

[8] V. Strotkötter, Y. Li, A. Kostka, F. Lourens, T. Löffler, W. Schuhmann, A. Ludwig, *Self-formation of compositionally complex surface oxides on high entropy alloys observed by accelerated atom probe tomography: a route to sustainable catalysts*, Materials Horizons 11 (2024) 4932–4941, 10.1039/d4mh00245h.

[9] Y.J. Li, A. Savan, A. Kostka, H.S. Stein, A. Ludwig, *Accelerated atomic-scale exploration of phase evolution in compositionally complex materials*, Materials Horizons 5 (2018) 86–92, 10.1039/C7MH00486A.

[10] B. Gault, A. Chiaramonti, O. Cojocaru-Mirédin, P. Stender, R. Dubosq, C. Freysoldt, S.K. Makineni et al., *Atom probe tomography*, Nature reviews. Methods primers 1 (2021), 10.1038/s43586-021-00047-w.

[11] M.P. Moody, A. Vella, S.S.A. Gerstl, P.A.J. Bagot, *Advances in atom probe tomography instrumentation: Implications for materials research*, MRS Bulletin 41 (2016) 40–45, 10.1557/mrs.2015.311.

[12] D. Larson, T. Prosa, R. Ulfig, B. Geiser, T. Kelly, *Local Electrode Atom Probe Tomography: A User's Guide*, 2013.

[13] O. Cojocaru-Mirédin, Y. Yu, J. Köttgen, T. Ghosh, C.-F. Schön, S. Han, C. Zhou et al., *Atom Probe Tomography: a Local Probe for Chemical Bonds in Solids*, Adv. Mater. 36 (2024) 2403046, 10.1002/adma.202403046.

[14] P. Gas, F.M. d'Heurle, *Formation of silicide thin films by solid state reaction*, Applied Surface Science 73 (1993) 153–161, 10.1016/0169-4332(93)90160-D.

[15] Y. Li, D. Zanders, M. Meischein, A. Devi, A. Ludwig, *Investigation of an atomic-layer-deposited Al2O3 diffusion barrier between Pt and Si for the use in atomic scale atom probe tomography studies on a combinatorial processing platform*, Surf Interface Anal 53 (2021) 727–733, 10.1002/sia.6955.

[16] S.M. Goodnick, M. Fathipour, D.L. Ellsworth, C.W. Wilmsen, *Effects of a thin SiO2 layer on the formation of metal–silicon contacts*, J. Vac. Sci. Technol. 18 (1981) 949–954, 10.1116/1.570962.

[17] S.J. Yoon, J.W. Jeon, J. Lee, J.T. Park, C. Lee, K.J. Yu, H. Bae et al., *Defect Formation and Electrical Transformation in SiO2 Thin Films via Ti-Induced Interdiffusion*, Acta Materialia 296 (2025) 121313, 10.1016/j.actamat.2025.121313.
[18] E. Akbarnejad, A. Kostka, Y. Li, M.K. Klein, K. Kozhuharov, G. Fritz, S. Kalt et al., *Enabling High-Temperature Atomic-Scale Investigations with Combinatorial Processing Platforms Using Improved Thermal SiO2 Diffusion and Reaction Barriers*, Adv. Mater. Interfaces 11 (2024) 2400138, 10.1002/admi.202400138.
[19] Y. Li, E. Akbarnejad, Q. Bizot, A. Kostka, N. Pukhareva, R. Zerdoumi, A. Savan et al., *A hidden low-temperature transformation pathway in compositionally complex materials*, arXiv preprint arXiv:2608.24357 (2026).
[20] K.S. Nakayama, M. Nishijima, Y. Zhang, C. Chen, M. Ueshima, K. Suganuma, *Metastable phases of Ag–Si: amorphous Si and Ag-nodule mediated bonding*, Scientific Reports 14 (2024) 19618, 10.1038/s41598-024-70298-6.
[21] T. Nakayama, S. Sotome, S. Shinji, *Stability and Schottky barrier of silicides: First-principles study*, Microelectronic Engineering 86 (2009) 1718–1721, 10.1016/j.mee.2009.03.018.
[22] C. Benazzouz, N. Benouattas, A. Bouabellou, *Competitive diffusion of gold and copper atoms in Cu/Au/Si and Au/Cu/Si annealed systems*, Nuclear Instruments and Methods in Physics Research Section B: Beam Interactions with Materials and Atoms 230 (2005) 571–576, 10.1016/j.nimb.2004.12.103.
[23] F Rollert, N A Stolwijk, H Mehrer, *Solubility, diffusion and thermodynamic properties of silver in silicon*, Journal of Physics D: Applied Physics 20 (1987) 1148, 10.1088/0022-3727/20/9/010.
[24] Z.M. Khumalo, C.T. Thethwayo, C.B. Mtshali, M. Msimanga, M.J. Madito, N. Numan, N. Mongwaketsi et al., *Interfacial reaction and phase formation in Pd/ZrO/Pd/TiO/Pd multilayer stack on silicon substrate: Investigated by ion beam techniques*, Vacuum 214 (2023) 112204, 10.1016/j.vacuum.2023.112204.
[25] A. Schrauwen, J. Demeulemeester, D. Deduytsche, W. Devulder, C. Detavernier, C.M. Comrie, K. Temst et al., *Ternary silicide formation from Ni-Pt, Ni-Pd and Pt-Pd alloys on Si(100): Nucleation and solid solubility of the monosilicides*, Acta Materialia 130 (2017) 19–27, 10.1016/j.actamat.2017.03.022.
[26] O. Abbes, K. Hoummada, D. Mangelinck, V. Carron, *Formation of Pt silicide on doped Si: Kinetics and stress*, Thin Solid Films 542 (2013) 174–179, 10.1016/j.tsf.2013.07.023.
[27] Y. Li, A. Kostka, A. Savan, A. Ludwig, *Investigation of Solid-State Interface Reactions Between the Compositionally Complex Solid Solution Noble Metal Alloy Ag-Au-Pd-Pt and Si*, Adv. Mater. Interfaces 12 (2025) 2400772, 10.1002/admi.202400772.
[28] Y.J. Li, A. Kostka, A. Savan, A. Ludwig, *Correlative chemical and structural investigations of accelerated phase evolution in a nanocrystalline high entropy alloy*, Scripta Materialia 183 (2020) 122–126, 10.1016/j.scriptamat.2020.03.016.
[29] M. Klinger, A. Jäger, *Crystallographic Tool Box (CrysTBox): automated tools for transmission electron microscopists and crystallographers*, Journal of applied crystallography 48 (2015) 2012–2018, 10.1107/S1600576715017252.